\documentclass[aps,pra,twocolumn,showpacs,nofootinbib,longbibliography,notitlepage]{revtex4-2}
\usepackage{etex}
\usepackage{amsmath,amssymb,amsthm}
\usepackage[colorlinks=true,citecolor=blue,urlcolor=blue]{hyperref}
\usepackage[pdftex]{graphicx}
\usepackage{times,txfonts}
\usepackage{braket}
\usepackage{color}
\usepackage{natbib}
\usepackage{amsmath,blkarray}
\usepackage{mathtools}
\usepackage{latexsym}
\usepackage{tabularx, booktabs}
\usepackage{graphics,epstopdf}
\usepackage{graphicx}
\usepackage{float}
\usepackage{graphicx}
\usepackage{amsfonts}
\usepackage{subcaption}
\usepackage{color,soul}

\newcommand{\be}{\begin{equation}}
\newcommand{\ee}{\end{equation}}
\newcommand{\ba}{\begin{eqnarray}}
\newcommand{\ea}{\end{eqnarray}}

\begin{document}
	\title{Characterizing  nonlocal correlations through various $n$-locality inequlities in quantum network  }
		\author{Sneha Munshi}
	\author{ A. K. Pan }
	\email{akp@nitp.ac.in}
	\affiliation{National Institute of Technology Patna, Ashok Rajpath, Patna, Bihar 800005, India}
	
	\begin{abstract}
The multipartite quantum networks feature multiple independent sources, in contrast to the conventional multipartite Bell experiment involving a single source. Despite the initial independence of resources, the multiple observers in the network can suitably choose measurements on their local subsystems and generate a form of quantum nonlocality across the network. So far, network nonlocality has been explored when each source produces a two-qubit entangled state. In this work, we demonstrate the network nonlocality when each party performs a black-box measurement, and the dimension of the system remains unspecified. In an interesting work, by considering each source produces two-qubit entangled states in the conventional bilocal scenario, Gisin \emph{et. al.} in [ \href{https://doi.org/10.1103/PhysRevA.96.020304} {Phys. Rev. A {\bf 96}, 020304 (2017)}]  demonstrated a correspondence between the violations of bipartite Clauser-Horne-Shimony-Halt inequality and the bilocality inequality. We introduce a variant of the sum-of-squares approach to reproduce their results without assuming the dimension of the system. We then generalize the argument for network nonlocality in star-network topology. Further, we propose a new set of $n$-locality inequalities in star-network configuration where each of the $n$ parties performs an arbitrary number of dichotomic measurements and demonstrate the above correspondence between the quantum violations of the $n$-locality inequalities and the chained Bell inequalities. A similar correspondence is demonstrated based on a recently formulated family of $n$-locality inequalities whose optimal quantum violation cannot be obtained when each source emits a two-qubit entangled state and requires multiple copies of two-qubit entangled states. Throughout this paper, each party in the network performs black-box measurements, and the dimension of the system remains unspecified.  
	
	\end{abstract}
	\maketitle
	\section{Introduction}
	
Bell's theorem \cite{Bell} states that the local hidden variable theory is incapable of reproducing all the predictions of quantum theory. This outstanding fundamental quantum mechanical feature is widely known as quantum nonlocality. Apart from the immense impression of Bell's theorem in quantum foundations, this theorem created a revolutionary impact in the practical implementation of quantum-enabled science and technology. Since the last two decades, it has influenced almost all the research of quantum foundation and information and heavily contributed to the development of quantum technology, especially in the modification of the security protocols of different communication complexity problems where quantum nonlocality was the central resource of enhancing the reliability of the protocol \cite{brunnerreview}.       
	
	A typical Bell experiment comprises two space-like separated parties, say, Alice and Bob who share a common state. Alice (Bob) randomly applies measurements on her (his) subsystem upon receiving inputs $x\in \{1,2\}(y\in \{1,2\})$ and returns outputs $a\in \{0,1\}(b\in \{0,1\})$. Classically, the shared state, commonly known as hidden variable $\lambda$, predicts the outcome of the measurement (reality), which is  independent of the choice of measurement and the outcome of the other party (locality). It is dispensable for $\lambda$ to be constant in all runs of the experiment, and thus, different values of $\lambda$ during different runs are characterized by the probability distribution $\rho(\lambda)$. The joint probability of the outcomes $a$ and $b$ can be written in the factorized form as
	\begin{equation}
	\label{jp}
	P(a,b|x,y)=\int \rho(\lambda)P(a|x,\lambda)P(b|y,\lambda)d\lambda
	\end{equation} 
	In quantum theory, Alice and Bob may share an appropriate entangled state. In such a case, for suitable choices of observables by Alice and Bob, the joint probability distribution $P(a,b|x,y)$ deviates from the factorized form of Eq.(\ref{jp}), thereby violating the notion of local realism. This feature is commonly demonstrated through suitably formulated Bell's inequalities whose quantum violation unequivocally establishes the quantum nonlocality \cite{brunnerreview, CHSH1969}.
	
Quantum nonlocality in the multiparty scenario is a straightforward generalization of bipartite Bell nonlocality, unfolding a more potent form of nonlocal correlations. Multipartite quantum nonlocality \cite{brunnerreview, Horodecki2009,Guhnea} has been extensively studied for the last two decades. In conventional multipartite Bell experiments, a single source distributes the entangled system to multiple distant parties.     In contrast, Branciard \emph{et. al.} \cite{Bran2010,Cyril2012} demonstrated a curious form of multiparty nonlocal correlations in networks featuring multiple independent sources. In particular, they considered a tripartite network involving two independent sources and proposed nonlinear bilocality inequalities, which are violated by quantum theory. Since then, different types of topological structures of quantum network are constructed \cite{Tavakoli2016,Tava2016,Frit2016,Rosse2016,Chav2016,Tava2017,Andr2017,Fras2018,Luo2018,Lee2018,Gupta2018,Cyril2019,Renou2019,Kerstjens2019,Aberg2020,Banerjee2020,Gisi2020,Supic2020,Tavakoliarxiv,Luo2020,kundu2020,Mukh2020,Mukh2019,Gisinarxiv2021,Kraftarxiv,scarani,cont,Tavareview,Pompili2021,Kozl2020} featuring multiple independent sources, each dispensing independent subsystems to some particular set of observers. Despite the initial independence of the sources,  the observers can suitably choose their measurements settings to demonstrate nonlocal correlations. 

One particularly interesting topology is the star-shaped network involving arbitrary $n$ number of edge parties, each sharing an independent subsystem with the central party, originating from $n$ independent sources \cite{Frit2012, Armi2014}. In a star-network, the quantum violation of some suitably constructed $n$-locality inequalities reveals non-$n$-nonlocality in a star network. Sometimes, complete independence of resources can be challenging to implement in experiments, and the level of independence is analyzed in a triangle network \cite{Supic2020}. Interestingly,  even without considering any input for the observers, the outcomes of the fixed measurement are capable of revealing nonlocality across the triangle network \cite{Renou2019}. In recent times, several experiments have been performed to test the quantum violations of bilocality and $n$-locality in networks \cite{Saunders2017,Andreoli2017,Carvacho2017,Sun2019,Poderini2020,Agresti2020}. 
	
	In the bilocal scenario in the linear network, Gisin \emph{et. al.} \cite{Gisin2017} found an interesting connection between the violations of Clauser-Horne-Shimony-Halt (CHSH) inequality and bilocality inequality. Using the work of Horodecki \emph{et al.} \cite{Horodecki2009} for any two-qubit entangled state, they proved that any quantum state that violates CHSH inequality also violates the bilocality inequality \cite{Cyril2012}. It is important to note here that the network nonlocality so far is demonstrated by assuming the dimension of the system. In particular, the optimal quantum violation has been derived considering each source produces a two-qubit entangled state. We go beyond the dimension-dependent scenario by introducing a variant of the sum-of-squares (SOS) approach that enables us to derive the optimal quantum violations of various network inequalities.  

Starting from the simplest bilocality scenario in the linear tripartite network, we first derive the quantum violation of bilocality inequality without assuming the dimension of the system. We then reproduce Gisin's \emph{et. al.} result \cite{Gisin2017} which was derived by considering that each source emits a  two-qubit entangled state. We extend our approach to the $n$-locality scenario in star-network configuration featuring an arbitrary $n$ number of independent sources, $n$  number of edge party (say, Alice$_k$, with $k=1,2,\ldots, n$) and one central party Bob.   We then generalize our treatment for a recently proposed family of $n$-locality inequalities \cite{Sneha2021} in the star-network scenario featuring $m$ number of measurements for each edge party and $2^{m-1}$ number of measurements by the central party. Further, we propose a new set of $n$-locality inequalities when each observer performs an arbitrary $m$ number of dichomatic measurements and establish a similar correspondence of the violations of the $n$-locality inequalities and the chained Bell inequalities. Specifically, we show  that if an entangled state $\rho_{A_kB}$, shared between Alice$_k$ and Bob, violates chained two-party  Bell's inequalities in \cite{Ghorai2018} then $\bigotimes\limits_{k=1}^n\rho_{A_kB}$ violates our proposed set  of $n$-locality inequalities \cite{Sneha2021}. Throughout this paper, we assume the black-box measurements and the dimension of the system remains unspecified. Optimal violations of various $n$-locality inequalities thus enable self-testing of state and measurements. 
	
 The plan of the paper is the following. In Sec. \ref{SecII}, we derive the optimal violation of well-known CHSH inequality using the SOS approach. In Sec. \ref{SecIII}, we sketch the conventional bilocality and $n$-locality scenarios. Also, we discuss Gisin's \emph{et. al.} \cite{Gisin2017} result using the Horodecki criteria.	In Sec. \ref{SecIV}, we reproduce the similar correspondence using the SOS approach for both bilocality and $n$-locality scenarios. In Sec.  \ref{SecV}, we extend our treatment for the generalized $n$-locality inequality in star-network configuration \cite{Sneha2021} and set up an association of these inequalities with a suitable Bell-type inequality \cite{Ghorai2018}. Then, in Sec.  \ref{SecVI} we discuss the quantum violation of chained  Bell inequality without assuming dimension. Considering the scenarios, when each party performs $m=3,4$  number of measurements, we provide a dimension-independent optimization using the SOS approach. Further, we formulate a suitable $n$-locality inequality, derive its optimal violation, and establish the aforementioned correspondence with the quantum violation of chained Bell inequality.   We summarize our results and propose some open questions in Sec. \ref{SecVII}.

\section{ CHSH inequality and it's optimal violation}
\label{SecII}
The optimal quantum violation of CHSH inequality is well known and derived by Tsirelson \cite{tsi}. For our purpose here, we provide an alternative derivation of it using a variant of the SOS approach \cite{pan21}. In CHSH scenario, two parties Alice and Bob, perform two dichotomic measurements $\{X_1,X_2\}$ and $\{Y_1,Y_2\}$ respectively. The  CHSH inequality is given by 
\ba
\label{chsh1}
\mathcal{B}=(X_1+X_2)Y_1+(X_1-X_2)\leq 2
\ea
which is valid for any classical theory satisfying locality and realism. The optimal quantum value of the Bell expression is $(\mathcal{B})_{Q}^{opt}=2\sqrt{2}$.

To derive $(\mathcal{B})_{Q}^{opt}$ using the SOS approach, consider that $(\mathcal{B})_{Q}\leq \delta$  where $\delta$ is the upper bound of $(\mathcal{B})_{Q}$.  Equivalently, one can assume that there is a positive semidefinite operator $\langle \gamma_{\mathcal{B}}\rangle\geq 0$ which can be expressed as $\langle \gamma_{\mathcal{B}}\rangle_{Q}=-(\mathcal{B})_{Q}+\delta$.	We define 
\ba
	\label{gammaxy}
	\gamma_{\mathcal{B}}=\dfrac{\omega'}{2}(M_{1})^{\dagger}M_1+ \frac{\omega''}{2}(M_{2})^{\dagger}M_{2}
	\ea 
where $M_{1}$ and $M_2$ are the suitable positive operators which are polynomial functions of   $X_1,X_2, Y_1$ and $Y_2$. Without loss of generality, we assume 

	\ba
	\label{mmxy}
M_{1}|\psi\rangle_{AB}=\bigg(\frac{{X}_{1}+{X}_{2}}{\omega'}\bigg)|\psi\rangle_{AB}
		 -Y_1|\psi\rangle_{AB}\\
		\label{mmxy1}
		 		M_{2}|\psi\rangle_{AB}=\bigg(\frac{{X}_{1}-{X}_{2}}{\omega''}\bigg)|\psi\rangle_{AB}-Y_2|\psi\rangle_{AB}
	\ea
	where $\omega'$ and  $\omega''$ are suitable positive numbers  defined as $\omega'=||(X_1+X_2)|\psi\rangle_{AB}||_2=\sqrt{2+\langle\{X_1,X_2\}\rangle},
	$ and $\omega''=||(X_1-X_2)|\psi\rangle_{AB}||_2=\sqrt{2-\langle\{X_1,X_2\}\rangle}$.

	Putting Eqs. (\ref{mmxy}) and (\ref{mmxy1}) in Eq. (\ref{gammaxy}), and further rearranging we get 	
	\begin{align}
		(\mathcal{B})_Q = \left(\omega'+\omega''\right) -\langle\gamma_{\mathcal{B}}\rangle_{Q}
	\end{align}
		Since $\langle \gamma_{\mathcal{B}}\rangle_{Q}\geq 0$, we have the optimal value of $ (\mathcal{B})_Q$ as
	\ba 
	\nonumber
	 (\mathcal{B})_Q^{opt}&=&max[ \omega'+\omega'']\\\nonumber
	 &=&max\left[\sqrt{2+\langle\{X_1,X_2\}\rangle}+\sqrt{2-\langle\{X_1,X_2\}\rangle}\right]
	\ea
	Clearly, the maximization requires $\{X_1,X_2\}=0$, which yields  $(\mathcal{B})_Q^{opt}=2\sqrt{2}$. Also, the optimal quantum value imply that the following conditions 
	\begin{align}
	\label{mm}
	M_{1}|\psi\rangle_{AB}=0;\hspace{2mm} 	M_{2}|\psi\rangle_{AB}=0
	\end{align}
has to be satisfied. This in turn fixes the Bob's observables and required entangled state for achieving optimal value.  Hence, optimal value $(\mathcal{B})_Q^{opt}$ certifies that both Alice's and Bob's observables to be mutually anticommuting and the shared state is maximally entangled.  
\section{Bilocality Scenario}
 \label{SecIII}
	\begin{figure}[h]
				\includegraphics[scale=0.4]{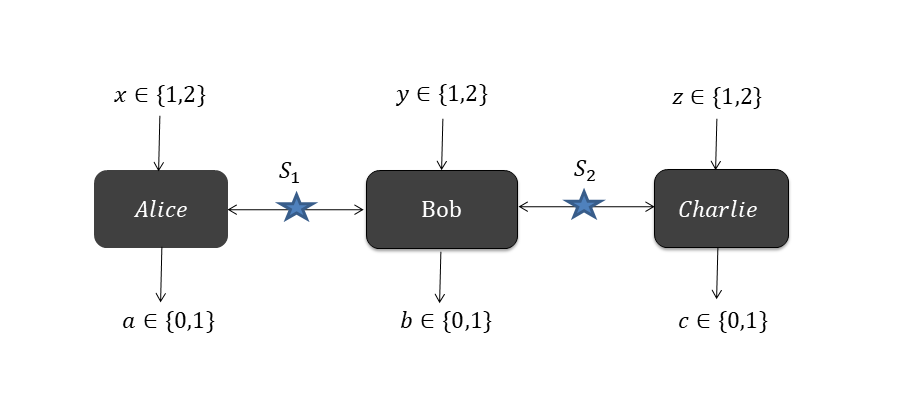}
				\caption{
				The source $S_1(S_2)$ shares system between Bob and Alice(Charlie) }
				\label{FIG1}
				\end{figure}

We start encapsulating the simplest non-trivial network - the bilocal scenario \cite{Cyril2012} which involves two edge parties and one central party, as depicted in Fig \ref{FIG1}. The source $S_{1}$ $(S_2)$ distribute systems to Alice (Charlie) and Bob. Alice and Charlie, each performs two binary-outcome measurements on their respective subsystems upon receiving inputs   $x,z\in\{1,2\}$, and return the outputs $a$, $c\in \{0,1\}$ respectively. Upon receiving inputs $y\in\{1,2\}$, Bob performs two binary outcome measurements on the system he receives from two sources. In classical hidden variable theory, one assumes that  the sources $S_1$ and $S_2$ distributes classical variables  $\lambda_1$ and $\lambda_2$ respectively.

The crucial assumption in the context of network is that  the sources $S_1$ and $S_2$ are independent to each other \cite{Cyril2012} and hence $\lambda_{1}$ and $\lambda_{2}$ are uncorrelated.  The joint distribution  $\rho{(\lambda_{1},\lambda_{2})}$ can then be factorized as  $
	\rho{(\lambda_{1},\lambda_{2})} = \rho_{1}{(\lambda_{1})}\rho_{2}{(\lambda_{2})}
$,  satisfying $\int d\lambda_{1}\rho_{1}{(\lambda_{1})}=1$ and $\int d\lambda_{2}\rho_{2}{(\lambda_{2})}=1$.  Then the  triple-wise joint probability in the tripartite network scenario satisfying the bilocality condition can be expressed as \cite{Cyril2012}
	\begin{eqnarray}
	\nonumber
	P(a,b,c|x,y,z)&=&\int\int d\lambda_{1} d\lambda_{2}\hspace{2mm}\rho_{1}(\lambda_{1})\hspace{1mm}\rho_{2}(\lambda_{2})\\
	&&\times P(a|x,\lambda_{1})P(b|y,\lambda_{1},\lambda_{2})P(c|z,\lambda_{2}).\hspace{9pt}
	\end{eqnarray}
	Clearly, the outcome of Alice and Charlie solely depends on $\lambda_{1}$ and $\lambda_{2}$ respectively. However outcome of  Bob is dependent on both $\lambda_{1}$ and $\lambda_{2}$.
	
	Assuming Alice's  observables are  $A_{1}$ and $A_{2}$, Bob's observables are $B_{1}$ and $B_{2}$, and Charlie's   observables are $C_{1}$ and $C_{2}$ and considering $P(a,b,c|x,y,z)$ satisfying the  bilocality condition, Branciard \emph{et. al.}\cite{Cyril2012} formulated the nonlinear bilocality inequality which 
	is given by 
	\ba
	\label{Delbl22}
	(\mathcal{S})_{bl}=\sqrt{|{I}_1|}+\sqrt{|{I}_2|}\leq {2}
	\ea
 where `$bl$' denotes bi-locality. Here, $I_1$ and $I_2$ are the linear combinations of suitably chosen correlations defined as 
	\begin{eqnarray}
	\label{bl21}
	I_ 1=\langle(A_{1}+A_{2}) B_{1} (C_{1}+C_{2})\rangle\\
	\label{bl22}
	I_2=\langle(A_{1}-A_{2}) B_{2} (C_{1}-C_{2})\rangle
	\end{eqnarray}
    	with  \ba\nonumber\langle{A_xB_yC_z}\rangle = \sum_{a,b,c}(-1)^{a+b+c}\hspace{1mm}P(a,b,c|x,y,z)\ea
	In quantum theory, one may consider that the sources $S_1$ and $S_2$ emit two entangled states $\rho_{AB}$ and $\rho_{BC}$. It has been  shown \cite{Cyril2012} that the optimal quantum value $(\mathcal{S})_{Q}^{opt}=2\sqrt{2} >(\mathcal{S})_{bl}$. For optimal value of $	(\mathcal{S})_{Q}$, the observables of Alice (Charlie) has to be anticommuting, and Bob's observables ${B}_{1}$ and ${B}_{2}$ are commuting. The shared quantum states $\rho_{AB}$ and $\rho_{BC}$ has to be two-qubit maximally entangled state. 
	
	It is important to note that the quantum network scenario has been studied so far by considering the case when each source emits a two-qubit entangled state. The quantity $(S)_{Q}$ can be optimized when the local system of Alice and Charlie are qubit system, and hence, it is enough for both the sources $S_1$ and $S_2$ to emit a two-qubit maximally entangled state. However, the device-independence scenario demands dimension independence of Hilbert space. Here we go beyond the dimension restriction and derive the optimal quantum value of $(S)_{Q}$ by using an elegant SOS approach. We then derive Gisin's \emph{et. al.}\cite {Gisin2017} result that demonstrates the correspondence between the violations of CHSH inequalities and non-bilocality. Specifically, Gisin \emph{et. al.}\cite{Gisin2017} provided an interesting characterization by showing that if the states  $\rho_{AB}$  and $\rho_{BC}$  violates CHSH inequality in Eq.(\ref{chsh1}) 
	then $\rho_{AB}\otimes \rho_{BC}$
violates bilocality inequality  in Eq.(\ref{Delbl22}).  We extend our treatment to the $n$-locality scenario involving an arbitrary number of parties. We  generalize our treatment for arbitrary $m$ input for each edge party in a $n$-locality scenario in star-network. Further, we proposed a new set of new $n$-locality inequalities in star-network topology and demonstrated that if $\rho_{A_kB}$   violates chained Bell inequality for each $k\in[n]$, then $\bigotimes\limits_{k=1}^n\rho_{A_kB}$ violates our proposed set of  $n$-locality inequalities.		Once again, throughout this paper, we assume black-box measurements and put no constraint on the dimension of the system.

\subsection{The results of Gisin \emph{et. al.}\cite{Gisin2017}  }
\label{A}
The criteria for any arbitrary two-qubit entangled state violating bilocality inequality $(\mathcal{S})_{bl}\leq 2$ was provided by Gisin \emph{et. al.}\cite{Gisin2017}. However,  the proof assumes two-qubit entangled state and is based on the work of  Horodecki \emph{et. al.} \cite{Horodecki2009}.

Let the sources $S_1$ and $S_2$ emitting  two-qubit states $\rho_{AB}$ and $\rho_{BC}$ respectively  which can be expressed in terms of Pauli basis as 
\ba
\rho_{AB}=\dfrac{1}{4}\bigg(\mathbb{I}+\vec{m_{A}}\cdot \sigma\otimes\mathbb{I}+\mathbb{I}\otimes\vec{m_B}\cdot \sigma+\sum\limits_{rs}t^{AB}_{rs}\sigma_{r}\otimes\sigma_{s}\bigg)\hspace{4mm}\\\nonumber
\rho_{BC}=\dfrac{1}{4}\bigg(\mathbb{I}+\vec{m_{C}}\cdot \sigma\otimes\mathbb{I}+\mathbb{I}\otimes\vec{m_B}\cdot \sigma+\sum\limits_{st}t^{BC}_{st}\sigma_{s}\otimes\sigma_{t}\bigg)\hspace{4mm}
\ea
 For $\rho_{AB}$, the vectors $\vec{m_{A}}$ and $\vec{m_B}$ represent the Bloch vectors of the reduced state of Alice and Bob, and $t^{AB}_{rs}$ are the  corresponding correlation matrix, where $r,s\in\{\hat{x},\hat{y},\hat{z}\}$, and similarly for $\rho_{BC}$.   Let the observables of Alice are represented by the Bloch vectors $\vec{a}$ and $\vec{a}'$ and the observables of Charlie are represented by the Bloch vectors $\vec{c}$ and $\vec{c}'$,   and Bob performs Bell state measurement. The  Eq. (\ref{bl21}) can then be written as
\ba
I_1&=&Tr[(\vec{a}+\vec{a}')\cdot \vec{\sigma}\otimes  \sigma_z\otimes  \sigma_z\otimes(\vec{c}+\vec{c}')\cdot\vec{\sigma}(\rho_{AB}\otimes \rho_{BC})  ]\hspace{1mm}\\
\nonumber
&=&\sum\limits_{r}(\vec{a_{r}}+\vec{a_{r}}')t_{rz}^{AB}\sum\limits_{s}(\vec{c_{s}}+\vec{c_{s}}')t_{zs}^{AB}
\ea
and similarly $I_{2}$ can also be written. The correlation matrix  $t^{AB}=U^{AB}R^{AB}$, where   $U^{AB}$ is the unitary matrix and $R^{AB}=\sqrt{(t^{AB})^{\dagger}t^{AB}}\geq 0$. $R^{AB}$ has  three eigen values $\alpha^{A}_1\geq\alpha^{A}_2\geq \alpha^{A}_3\geq 0 $. Similarly, for $R^{BC}=(t^{BC})^{\dagger}t^{BC}$,  the corresponding eigen values are $\eta^{C}_1\geq\eta^{C}_2\geq \eta^{C}_3\geq 0 $. Since Bob is perfroming the measurements on the two-qubit system, hence, its observers are defined along $\hat{z}$ and $\hat{x}$ directions of Bloch sphere. Here, the $\hat{z}$ and $\hat{x}$ Bob uses with Alice can be different that $\hat{z}$ and $\hat{x}$ he uses with Charlie. The expression  $(\mathcal{S})_{Q}$ is  maximized  with respect to Bloch vectors $\vec{a},\vec{a'},\vec{c}$ and $\vec{c'}$. Clearly, they should lie in the two-dimensional subspace spanned by the eigenvectors of the largest two  eigen values \cite{Horodecki2009}. Hence, assume $\vec{a}=(\sin{\theta_1},0,\cos{\theta_1}),\vec{a}'=(\sin{\theta'_1},0,\cos{\theta'_1}), \vec{c}=(\sin{\theta_2},0,\cos{\theta_2}),\vec{c}'=(\sin{\theta'_2},0,\cos{\theta'_2})$ and maximizing with respect to $\theta_1,\theta'_1,\theta_2,\theta'_2$, one gets $\theta'_1=-\theta_1, \theta'_2=-\theta_2 $. Using this in the expression of $(\mathcal{S})_Q$, one gets $(\mathcal{S})_{Q}^{opt}=2\sqrt{\alpha^{A}_1\eta^{C}_1+\alpha^{A}_2\eta^{C}_2}$.  Using Horodecki's criteria \cite{Horodecki2009}, we get  the maximal violation of CHSH inequality for  the state $\rho_{AB}$ is given by $\mathcal{B}^{max}_{AB}=2\sqrt{(\alpha^{A}_1)^2+(\alpha^{A}_2)^2}$ and for  $\rho_{BC}$, $\mathcal{B}^{max}_{BC}=2\sqrt{(\eta^{C}_1)^2+(\eta^{C}_2)^2}$. We can then write

\ba\label{Gsq}(S)_{Q}^{max}\leq \sqrt{\mathcal{B}^{max}_{AB}\cdot \mathcal{B}^{max}_{BC}}\ea
This result implies that if $\rho_{AB}$ or $\rho_{BC}$ violate  $\mathcal{B}_{AB}\leq 2$ and  $\mathcal{B}_{BC}\leq 2$ respectively, then $\rho_{AB}\otimes \rho_{BC}$ violates the   bilocality inequality in Eq. (\ref{Delbl22}). This proof is for any arbitrary given two-qubit entangled state. Here we will provide a proof that is independent of the dimension of the system.
\section{An elegant and dimension-independent   derivation of  Gisin's work}
\label{SecIV}
As mentioned, the above proof is based on   Horodecki \emph{et. al.} \cite{Horodecki2009} that  uses two-qubit entangled state. To go beyond this dimensional  restriction, we introduce a variant of SOS approach first initiated in \cite{AKP2020} to establish Gisin \emph{et. al.}'s \cite{Gisin2017} claim for arbitrary dimensional entangled state. In doing so, let us consider that $(\mathcal{S})_{Q}\leq \beta$  where $\beta$ is clearly the upper bound of $(\mathcal{S})_{Q}$.  This is equivalent to showing that there is a positive semidefinite operator $\langle \gamma_S\rangle_{Q}\geq 0$ which can be expressed as $\langle \gamma_S\rangle=-(\mathcal{S})_{Q}+\beta$.	This can be proven by considering a set of suitable positive operators $M'_{1}$ and $M'_2$ which are polynomial functions of   $A_{x}$ , $B_{y}$ and $C_z$, so that,
	\begin{align}
	\label{gamma1}
	\gamma_S=\dfrac{\sqrt{\omega_{1}}}{2}(M'_{1})^{\dagger}M'_1+ \frac{\sqrt{\omega_{2}}}{2}(M'_{2})^{\dagger}M'_{2}
	\end{align} where $\omega_1=(\omega_{1})_{A}\cdot (\omega_{2})_{C}$ are suitable positive numbers that  will be specified soon. 
	For our purpose, we suitably choose the  positive operators $M'_{1}$ and $M'_{2}$ as
	\begin{equation}
	\label{mABC}
	\begin{split}
M'_{1}|\psi\rangle_{ABC}=\sqrt{\bigg|\left(\frac{{A}_{1}+{A}_{2}}{(\omega_{1})_{A}}
		\otimes\frac{C_{1}+C_{2}}{(\omega_{1})_{C}}\right)|\psi\rangle_{ABC}
		\bigg|} -\sqrt{| B_1|\psi\rangle_{ABC}|}\\
M'_{2}|\psi\rangle_{ABC}=\sqrt{\bigg|\left(\frac{{A}_{1}-{A}_{2}}{(\omega_{2})_{A}}
		\otimes\frac{C_{1}-C_{2}}{(\omega_{2})_{C}}\right)|\psi\rangle_{ABC}
		\bigg|} -\sqrt{| B_2|\psi\rangle_{ABC}|}
		\end{split}
	\end{equation}
	where $(\omega_{1})_{A}=||({A}_{1}+{A}_{2})|\psi\rangle_{AB}||_{2}=\sqrt{2+\langle\{A_{1},A_{2}\}\rangle}$. Similarly, \hspace{3pt} $(\omega_{1})_{C}=\sqrt{2+\langle\{C_{1},C_{2}\}\rangle}$ \hspace{2pt} and              $(\omega_{2})_{A}=\sqrt{2-\langle\{A_{1},A_{2}\}\rangle}$, \hspace{3pt} $(\omega_{2})_{C}=\sqrt{2-\langle\{C_{1},C_{2}\}\rangle}$.
	Putting $M'_1$ and $M'_2$ from Eq. (\ref{mABC}) in Eq. (\ref{gamma1}), we get 	$\langle\gamma_S\rangle_{Q}=-(\mathcal{S}_{Q})         +\left(\sqrt{\omega_{1}}+\sqrt{\omega_{2}}\right)$. 	The optimal value of $(\mathcal{S})_{Q}$ is obtained if  $\langle \gamma_S\rangle_{Q}=0$. This in turn provides,
	
	\begin{eqnarray}
	\label{SQopt}
	\nonumber(\mathcal{S})_{Q}^{opt}&=& max[\sqrt{\omega_{1}}+\sqrt{\omega_{2}}]\\
	&=&max[\sqrt{(\omega_{1})_{A}\cdot(\omega_{1})_{C}}+\sqrt{(\omega_{2})_{A}\cdot(\omega_{2})_{C} }]
	\end{eqnarray}
 The condition $\langle \gamma_S\rangle_{Q}=0$ yields 
	\begin{align}
	\label{mm}
	M'_{1}|\psi\rangle_{ABC}=0;\hspace{2mm} 	M'_{2}|\psi\rangle_{ABC}=0
	\end{align} which provides the form of $B_1$ and $B_2$ required for optimization.  Now, if we consider the  source $S_1(S_2)$ emits  the entangled state, shared  between  Alice ( Charlie) and Bob is $\rho_{AB}(\rho_{BC})$,  and  Bob performs measurement $B_1$ and $B_2$ on his part of the subsystem  then the  relevant CHSH  inequality (as given in Eq. (\ref{chsh1})) $\mathcal{B}_{AB} (\mathcal{B}_{BC}$)   will be optimized using the similar SOS approach as stated in Sec. \ref{SecII}, and  we then have,
	\ba
	 (\mathcal{B}_{AB})^{max}=max[(\omega_1)_{A}+(\omega_2)_{A}]\\
	 \nonumber
	 (\mathcal{B}_{BC})^{max}=max[(\omega_1)_{C}+(\omega_2)_{C}]
	\ea
	
	Using the inequality, 
	$\sqrt{r_{1}s_{1}}+\sqrt{r_{2}s_{2}}\leq\sqrt{r_{1}+r_{2}}\sqrt{s_{1}+s_{2}}$ for $r_{1}, s_{1}, r_{2}, s_{2}\geq 0$, we can write Eq.(\ref{SQopt}) as
	\begin{equation}
	\label{dd}
	(\mathcal{S})^{opt}_{Q}\leq\bigg(\sqrt{(\omega_1)_{A}+(\omega_2)_{A}}\sqrt{(\omega_1)_{C}+(\omega_2)_{C}}\bigg)
	\end{equation} 
which in turn provides 
	\begin{equation}
	\label{dd}
	(\mathcal{S})^{opt}_{Q}\leq\ \sqrt{\mathcal{B}_{AB}^{opt}\cdot\mathcal{B}_{BC}^{opt}}
	\end{equation}
	
	This is precisely the result in Eq.(\ref{Gsq}). However, our derivation does not assume the dimension of the system in contrast to Gisin \emph{et. al.}'s \cite{Gisin2017} derivation.
	
	
	\subsection{Generalization for $n$-locality in  star-network configuration}
	We further generalize the above  bilocality scenario  to  $n$-locality scenario in star-network configuration, which involves $n$ independent sources and $n+1$ number of  independent observers. There are  $n$ edge observers (Alices) and one central observer (Bob). Each independent source S$_{k}$  shares a physical system with one edge party Alice$_{k}$ and the central party Bob,  where $k\in[n]$. Let,  the observables of $k^{th}$ Alice are denoted by $A^{k}_{1}$ and $A^{k}_{2}$,  and Bob's  observables are $B_{1}$ and $ B_{2}$. The $n$-locality inequality can be defined as \cite{Armi2014}  \begin{equation}
	\label{Sn}
		(\mathcal{S}^{n})_{nl}=|{I}^{n}_{1}|^{1/n}+|{I}^{n}_{2}|^{1/n}\leq {2}
		\end{equation} where '$nl$' denotes $n$-locality,  and   $I^{n}_{1}$ and $I^{n}_{2}$ are  defined as 
\begin{eqnarray*}
I^{n}_{1}=\langle\prod\limits_{k=1}^n(A^{k}_{1}+A^{k}_{2})B_{1}\rangle, 
I^{n}_{2}=\langle\prod\limits_{k=1}^n(A^{k}_{1}-A^{k}_{2})B_{2}\rangle
\end{eqnarray*}
Here, $B_1=\bigotimes\limits_{k=1}^n B_1^k$, $B_2=\bigotimes\limits_{k=1}^n B_2^k$ and $k$ corresponds to the subsystem received from the source $S_k$.
The optimal quantum value of $(S^n)_Q$ is  \ba(\mathcal{S}^{n})_{Q}^{opt}=2\sqrt{2}\ea i.e., same as bilocality violation \cite{Andr2017}.
	
 Using the similar SOS approach as stated above, we can express the optimal quantum value $(\mathcal{S}^{n})_{Q}^{opt}$ as
 \ba\label{Snq}(\mathcal{S}^{n})_{Q}^{opt}=max[{(\omega^{n}_{1})}^{1/n}+{(\omega^{n}_{2})}^{1/n}]
 \ea
  where $\omega^{n}_{1}=\prod\limits_{k=1}^{n}(\omega^{n}_{1})_{A_k}$. By using the inequality
	\begin{equation}
	\label{Tavakoli}
	\ \forall\ \ z_{k}^{i} \geq 0; \ \ \ \sum\limits_{i=1}^{2^{m-1}}\bigg(\prod\limits_{k=1}^{n}z_{k}^{i}\bigg)^{\frac{1}{n}}\leq \prod \limits_{k=1}^{n}\bigg(\sum\limits_{i=1}^{2^{m-1}}z_{k}^{i}\bigg)^{\frac{1}{n}}
	\end{equation}  Eq.(\ref{Snq}) can  be written as
	\ba(\mathcal{S}^{n})_{Q}&\leq&\prod \limits_{k=1}^{n} \big((\omega^{n}_{1})_{A_k}+(\omega^{n}_{2})_{A_k}\big)^{1/n}\ea

  If we consider the entangled state  generated by a source $S_k$, shared between  Alice$_k$ and Bob and the corresponding observables are $A_1^k, A_2^k$ and $ B_1, B_2$, respectively,  then the  CHSH inequality for every $k^{th}$ Alice and Bob can be written as  
 \ba \label{BAkB}\mathcal{B}_{k}=\langle(A^k_1+A^k_2)B^k_1\rangle+\langle(A^k_1-A^k_2)B^k_2\rangle\ea Since $k\in[n]$, there will be $n$ number of CHSH inequalities. The scenario is as if Bob independently performs Bell's inequality violation with each Alice on the physical system emitted from $S_k$. By following the  derivation in Sec. \ref{SecII}, for every $k$, we can then get 
 \ba \label{Bkopt}(\mathcal{B}_{k})^{opt}_Q=\max[(\omega^{n}_{1})_{A_k}+(\omega^{n}_{2})_{A_k}]\ea

Now, by using Eq.(\ref{Bkopt}), we can write\ba
	(\mathcal{S}^{n})_{Q}&\leq&\prod \limits_{k=1}^{n}[(\mathcal{B}_{A_kB})_Q]^{(1/n)}
 \ea
	Hence if for each $k\in[n]$, the state $\rho_{A_kB}$ violates CHSH inequality in Eq.(\ref{Bkopt}) and each source $S_k$ shares the state $\rho_{A_kB}$ between Alice$_k$ and Bob ($\forall k\in[n]$), then $\bigotimes\limits_{k=1}^n\rho_{A_kB}$ violates $n$-locality inequality given in Eq. (\ref{Sn}). We note again that the dimension of the system is not assumed in the above derivation.\\
	\section{ Further generalization for arbitrary inputs }
		\label{SecV}
		We now show the similar correspondence between the quantum violations of a suitably formulated $n$-locality inequality \cite{Sneha2021} and a form of Bell's inequality. In \cite{Ghorai2018}, a Bell's inequality was proposed 
		\ba
\label{nbell}
\mathcal{G}_{m} =\sum\limits_{x=1}^{m}\sum\limits_{i=1}^{2^{m-1}} (-1)^{y_x^{i}}A_{x}\otimes B_{i}\leq m\binom{m}{\lfloor{\frac{m-1}{2}}\rfloor}
\ea 
where $m$ is arbitrary. This inequality in fact arises while analyzing the $n$-bit random access code \cite{Ambainis,Ghorai2018}. It was shown that the success probability is solely dependent on the quantum violation of a Bell's inequality \cite{Ambainis,AKP2020,Asmita}. For our purpose here we independently use this inequality Eq. (\ref{nbell}) which has no immediate connection to the random access code communication game. 
Here, two parties Alice and Bob performs the measurements of  $m$ and $2^{m-1}$ dichotomic observables and produce outputs $a,b\in\{0,1\}$ respectively. The term 
	 ${y_{x}^{i}}$ takes value either $0$ or $1$ which is   fixed by using the encoding scheme used in \cite{Ambainis,Ghorai2018,AKP2020,Asmita}. Then for a given $i$, ${y_{x}^{i}}$  will fix the  values of $(-1)^{y_x^{i}}$ as $1$ or $-1$ in Eq. (\ref{nbell}). For this, let us  assume a set of bit string   $y^{\delta}\in \{0,1\}^{m}$ with $\delta\in \{1,2...2^{m}\}$. Every element of the bit string can be expressed as $y^{\delta}=y^{\delta}_{x=1} y^{\delta}_{x=2} y^{\delta}_{x=3} .... y^{{\delta}_{x}=m}$. For  example, if $y^{\delta} = 010...01$ then $y^{{\delta}}_{x=1} =0$, $y^{{\delta}}_{x=2} =1$, $y^{{\delta}}_{x=3} =0$, $\cdots y^{\delta}_{x=m} =1$  . Now, we denote  the length $m$ binary strings as $y^{i}$ those have $0$ as the first bit in $y^{\delta}$ and hence $i\in \{1,2...2^{m-1}\}$, the inputs for Bob. If $i=1$, all bits are zero in  the string $y^{1}$ which leads to $(-1)^{x^{i}}=1$ for every $x\in \{1,2 \cdots m\}$. \\
	 
	 It has been shown in \cite{Ghorai2018} that  the quantum value of $\mathcal{G}_{m}$ is
\ba
\label{pncbound}
(\mathcal{G}_{m})_{Q}\leq 2^{m-1}\sqrt{m}
\ea
		  For the  $n$-locality scenario in star-network configuration,  we consider that each of the $n$ number of Alices shares a state with Bob, generated by the independent sources $S_{k}$ with $k\in [n]$, as depicted in Fig. \ref{FIG. 2}.  Alice$_k$ performs $m$ number of binary-outcome measurements $A^{k}_{x_{k}}$ ($\forall x_{k}\in[m]$ , for any $k$). Bob receives fixed number of inputs $i \in \{1,2, \cdots,2^{m-1}\}$ and performs binary-outcome measurements on $n$ number of systems he receives from $n$ independent sources.
		  A form of $n$-locality   inequality  was introduced in  \cite{Sneha2021} as \begin{eqnarray}
	\label{Deltanmbl}\label{deltanmnl}
	(\Delta^{n}_{m})_{nl}=\sum\limits_{i=1}^{2^{m-1}}|I^{n}_{m,i}|^{\frac{1}{n}}\leq\sum\limits_{j=0}^{\lfloor\frac{m}{2}\rfloor}\binom{m}{j}(m-2j) 
	\end{eqnarray} 
	
	where $I^{n}_{m,i}$ for given $i$, $n$ and $m$ is given by 
	\begin{eqnarray}
	\label{jnmi}
	I^{n}_{m,i}= \prod\limits_{k=1}^{n}\bigg[\sum\limits_{x_{k}=1}^{m}(-1)^{y^i_{x_{k}}} A^{k}_{x_{k}}\bigg]\hspace{1pt}B_{i}
	\end{eqnarray}

	Again ${y_{x_{k}}^{i}}$ takes value either $0$ or $1$.   The values of ${y_{x_{k}}^{i}}$ is fixed using the similar method as described above  for the construction of Eq.(\ref{nbell}). 
It is interesting to note that the upper bound in Eq.(\ref{Deltanmbl}) and Eq.(\ref{nbell}) look different, but they are the \emph{same} value,  derived by entirely different  ways in two different contexts.

	 In order to find the optimal quantum value of the expression $(\Delta_{m}^{n})_{nl}$ \cite{Sneha2021}, we follow the similar SOS approach. Considering suitable operator $\langle\gamma^{n}_{m}\rangle\geq 0$, we find $\langle\gamma^{n}_{m}\rangle_Q=-(\Delta^{n}_{m})_{Q}+\left(\sum\limits_{i=1}^{2^{m-1}}(\omega^{n}_{m,i})^{\frac{1}{n}}\right)$. Hence 	\begin{align}
	\label{delnopt}
	(\Delta^{n}_{m})_{Q}^{opt}= max\left(\sum\limits_{i=1}^{2^{m-1}}(\omega^{n}_{m,i})^{\frac{1}{n}}\right)
	\end{align} where $\omega^{n}_{m,i}=\prod\limits_{k=1}^{n}(\omega^{n}_{m,i})_{A_k}$ and $(\omega^{n}_{m,i})_{A_k}=||\sum\limits_{x_{k}=1}^{m}(-1)^{y^{i}_{x_{k}}} A_{x_{k}}||_2$.  The detailed calculation is given in \cite{Sneha2021}.\\

	If we consider the entangled systems generated by source $S_k$  between  Alice$_k$ and Bob and the corresponding observables are $A_{x_k}^k$ and   $B_i $ ($x_k\in[m]$ , $i\in[2^{m-1}])$, then  the similar form of   Bell expression as in (\ref{nbell})   can be written as  
 \be \label{Gmk}(\mathcal{G}_{m})_k=\sum\limits_{i=1}^{2^{m-1}}\bigg\langle\sum\limits_{x_{k}=1}^{m}(-1)^{y^{i}_{x_{k}}} A_{x_{k}}^{k}B_i\bigg\rangle, \forall k\in[n]
 \ee
 To optimize  $[(\mathcal{G}_m)_k]_Q$ from Eq.(\ref{Gmk}),  we consider $\langle(\gamma^{n}_{m})_k\rangle_Q=[(\mathcal{G}_m)_k]_Q+(\zeta^{n}_m)_k$ where the lower index $k$  denotes the system generated by source $S_k$. Let 
 	\ba
	\label{gammanmk}
	\langle(\gamma^{n}_{m})_k\rangle=\sum\limits_{i=1}^{2^{m-1}}  \frac{{(\omega^{n}_{m,i}})_{A_k}}{2 }\langle\psi|(M^{n,k}_{m,i})^{\dagger}M^{n,k}_{m,i}|\psi\rangle
	\ea
	 $(\omega^{n}_{m,i})_{A_k}$ is a positive number such that $(\omega^{n}_{m,i})_{A_k}=||\sum\limits_{x_{k}=1}^{m}(-1)^{y^{i}_{x_{k}}} A_{x_{k}}^{k}||_2$. The optimal quantum value of $[(\mathcal{B}_m)_k]_Q$  will be  obtained when $\langle (\gamma_{m}^{n})_k\rangle_{Q}=0$, i.e., 
	\begin{align}
	\label{mmnm}
	\forall i, \ \ \ M^{n,k}_{m,i}|\psi\rangle_{A_kB}=0
	\end{align}
where  $|\psi\rangle_{A_{k}B}$ is originated from the independent source $S_{k}$, $\forall k\in[n]$. We suitably choose the operators $M^{n,k}_{m,i}$ as 
	\begin{eqnarray}
	\nonumber
	M^{n,k}_{m,i}|\psi\rangle_{A_kB}&=&\frac{1}{(\omega^{n}_{m,i})_{A_k}}\bigg[\sum\limits_{x_{k}=1}^{m}(-1)^{y^{i}_{x_{k}}} A_{x_{k}}^{k}\hspace{1pt}\bigg]|\psi\rangle_{A_kB}-B_{i}|\psi\rangle_{A_kB}\\\nonumber&=&\frac{1}{(\omega^{n}_{m,i})_{A_k}}I^{n,k}_{m,i}|\psi\rangle_{A_kB}-B_{i}|\psi\rangle_{A_kB}
	\end{eqnarray}
Here, for notational convenience,  we  write  $\bigg[\sum\limits_{x_{k}=1}^{m}(-1)^{y^{i}_{x_{k}}} A_{x_{k}}^{k}\hspace{1pt}\bigg]=I^{n,k}_{m,i}$. Hence,
	\begin{align}
	\label{Cymi}
			(\omega^{n}_{m,i})_{A_k}^{2}=\langle\psi| (I_{m,i}^{n,k})^{\dagger}(I_{m,i}^{n,k})|\psi\rangle= \langle\psi| (I_{m,i}^{n,k})^{2}|\psi\rangle.
	\end{align}
	Putting this in Eq. (\ref{gammanmk}) and using the facts that each observables, $A^k_{x_k}$ and $B_i$ are dichotomic, we get

     \be
     \langle(\gamma^{n}_{m})_k\rangle_{Q}=max\sum\limits_{i=1}^{2^{m-1}}(\omega^{n}_{m,i})_{A_k}-[(\mathcal{G}_m)_k]_{Q}\ee
	
	Since $(\gamma^{n}_{m})_k$ is positive semi-definite, the maximum value of $	[(\mathcal{G}_{m})_k]_{Q}$ is obtained when $\langle(\gamma^{n}_{m})_k\rangle_{Q}= 0$ i.e., 

		\be
		[(\mathcal{G}_m)_k]_{Q}^{opt}=max\sum\limits_{i=1}^{2^{m-1}}(\omega^{n}_{m,i})_{A_k}
		\ee
To obtain the optimal violation, the observables of each Alice$_k$ must be anti-commuting, and the state must be maximally entangled. It can be found in \cite{Ghorai2018, Sneha2021} that $[(\mathcal{G}_m)_k]_{Q}\leq 2^{m-1}\sqrt{m}$.

 Now, using the inequality  (\ref{Tavakoli}), from Eq.(\ref{delnopt}), we can write 
	\be
	(\Delta_{m}^{n})_{Q}\leq\prod \limits_{k=1}^{n} \bigg(\sum\limits_{i=1}^{2^{m-1}}(\omega^{n}_{m,i})_{A_k}\bigg)^{\frac{1}{n}}=\prod\limits_{k=1}^{n}\bigg([(\mathcal{G}_m)_k]_{Q}\bigg)^{\frac{1}{n}}\ee
	Hence, if each source  $S_k, (k\in[n])$ is emitting the entangled state $\rho_{A_k B}$ that violates the Bell's inequality  in Eq.(\ref{Gmk}) then $\bigotimes\limits_{k=1}^n\rho_{A_kB}$ assuredly violates the $n$-locality inequality in Eq.(\ref{deltanmnl}).
	
 \section{Correspondence between the violations of $n$-locality  and chained bell inequality}
\label{SecVI} 

In this section, we propose a new set of $n-$locality inequalities. Considering a similar $n$-locality scenario in star network configuration again, here we derive the aforementioned correspondence between the quantum violations of chained Bell inequality \cite{Brau1989} and a suitably constructed $n$-locality inequality. For chained  Bell inequality with an arbitrary $m$ number of measurements per party, we consider that Alice and Bob perform dichotomic measurements denoted by $A_i$ and $B_i$ with  $i\in[m]$. The chained Bell  inequality \cite{Brau1989} can  be written as 
 \ba
 \label{BmAB}
 (\mathcal{
C}_{m})_{AB}=\sum\limits_{i=1}^{m}(A_i+A_{i+1})B_i\leq {2m-2}
 \ea where $A_{m+1}=-A_1$.  Here we provide explicit derivations of quantum violation of chained Bell inequality for $m=3$ and $4$. The optimal violation of chained Bell inequality is commonly derived for a two-qubit entangled state. In our derivation, the dimension of the system is not bounded, and hence the optimal quantum violation of chained Bell inequality can be used for device-independent certifications. Such a derivation for chained Bell inequality has not hitherto been provided. 
\subsection{Chained Bell inequality for $m=3$ and $4$}
\label{SECVA}
Consider the scenario when Alice and Bob, each measuring three dichotomic observables  $A_i,$ and $B_i$ respectively with $i\in[3]$. The  relevant chained Bell inequality can be written as
    					\ba
    					(\mathcal{C}_3)_{AB}=(A_1+A_2)B_1+(A_2+A_3)B_2+(A_3-A_1)B_3\leq 4
    				\hspace{5mm}	\ea
    					
    				To optimize $(\mathcal{C}_3)_{AB}$, by following our SOS approach developed in Sec.  \ref{SecII}, we define a positive semidefinite operator $\langle \gamma_{\mathcal{C}_3}\rangle_{Q}\geq 0$ so that  $\langle \gamma_{\mathcal{C}_3}\rangle_Q=-[(\mathcal{C}_3)_{AB}]_{Q}+\beta_{\mathcal{C}_3}$ where $\beta_{\mathcal{C}_3}$ is a positive quantity.	By considering suitable positive operators $L_{3,i}$, where $ i\in[3]$ we can write
	\begin{align}
	\label{gammach3}
	\gamma_{\mathcal{C}_3}=\dfrac{\nu_{3,1}}{2}(L_{3,1})^{\dagger}L_{3,1}+ \dfrac{\nu_{3,2}}{2}(L_{3,2})^{\dagger}L_{3,2}+\dfrac{\nu_{3,3}}{2}(L_{3,3})^{\dagger}L_{3,3}
	\end{align} where $\nu_{3,i}$s are suitable positive numbers.
	We choose  $L_{3,i}$s as
	\begin{equation}
	\label{mmch3}
	\begin{split}
L_{3,1}|\psi\rangle_{AB}=\left(\frac{{A}_{1}+{A}_{2}}{\nu_{3,1}}\right)|\psi\rangle_{AB}
		 -| B_1|\psi\rangle_{AB}\\
L_{3,2}|\psi\rangle_{AB}=\left(\frac{{A}_{2}+{A}_{3}}{\nu_{3,2}}\right)|\psi\rangle_{AB}
		 -| B_2|\psi\rangle_{AB}\\
		 L_{3,3}|\psi\rangle_{AB}=\left(\frac{{A}_{3}-{A}_{1}}{\nu_{3,3}}\right)|\psi\rangle_{AB}
		 -| B_3|\psi\rangle_{AB}\\
		\end{split}
	\end{equation}
	where $\nu_{3,1}=||({A}_{1}+{A}_{2})|\psi\rangle_{AB}||_{2}=\sqrt{2+\langle\{A_{1},A_{2}\}\rangle}$. Similarly, \hspace{3pt} $\nu_{3,2}=\sqrt{2+\langle\{A_{2},A_{3}\}\rangle}$ \hspace{2pt} and              $\nu_{3,3}=\sqrt{2-\langle\{A_{1},A_{3}\}\rangle}$.
	Putting $L_{3,i}$s  from Eq. (\ref{mmch3}) in Eq. (\ref{gammach3}), we get 	
	\begin{align}
		[(\mathcal{C}_3)_{AB}]_{Q}=- \langle\gamma_{\mathcal{C}_3}\rangle_{Q}        +\left(\nu_{3,1}+\nu_{3,2}+\nu_{3,3}\right). 	
	\end{align}

Hence	the optimal value of $[(\mathcal{C}_3)_{AB}]_{Q}  $ is obtained when  $\langle \gamma_{\mathcal{C}_3}\rangle_{Q}=0$, i.e., \ba[(\mathcal{
C}_{3})_{AB}]_{Q}^{ opt}=max[(\nu_{3,1})+(\nu_{3,2})+(\nu_{3,3})]\ea	
Using the inequality  $\sum\limits_{i=1}^{n}f_i\leq \sqrt{n\sum\limits_{i=1}^{n}f_i^2}$, ($\forall f_i\geq 0 $) we can write 
 \ba\nonumber
 [(\mathcal{
C}_{3})_{AB}]_{Q}&\leq&\sqrt{3\big[(\nu_{3,1})^2+(\nu_{3,2})^2+(\nu_{3,3})^2\big]}\\
 \nonumber
 &=&\sqrt{3\big[6+\langle \{A_2,(A_1+A_3)\}\rangle-\langle\{A_1,A_3)\}\rangle\big]}
 \ea
 Considering $A_2=(A_1+A_3)/\nu'_3$, we get
 \ba[(\mathcal{
C}_{3})_{AB}]_{Q}\leq\sqrt{3\big[6+2\sqrt{2+\langle\{A_1,A_3)\}\rangle}-\langle\{A_1,A_3)\}\rangle\big]}\ea
 A simple calculation gives the maximization condition  $\{A_1,A_3)\}=-1$ which implies $\nu'_3=\sqrt{2+\{A_1,A_3)\}}=1$. Thus,  we get the condition on Alice's observables  $A_1-A_2+A_3=0$. Also, we have found $\{A_1,A_2\}=1$ and $\{A_2,A_3\}=1$ and consequently  $\nu_{3,1}=\nu_{3,2}=\nu_{3,3}=\sqrt{3}$. Bob's observables and the state required for this optimization can also be found from the condition $\langle\gamma_{\mathcal{C}_3}\rangle_Q=0$.   Finally,  we get  \ba[(\mathcal{C}_{3})_{AB}]_{Q}^{opt}=3\sqrt{3}\ea

Similarly, if we consider the scenario where each party performs four dichotomic observables $A_i,B_i,$ with $i\in[4]$ then, the chained Bell inequality for $m=4$  is
\ba
[(\mathcal{
C}_{4})_{AB}]&=&(A_1+A_2)B_1+(A_2+A_3)B_2\\
\nonumber
&+&(A_3+A_2)B_3+(A_4-A_1)B_4\leq 6\hspace{0.6cm}
 \ea
Following the similar SOS approach, we get 	\begin{eqnarray}[(\mathcal{
C}_{4})_{AB}]_{Q}^{opt}&=& max\left[(\nu_{4,1})+(\nu_{4,2})+(\nu_{4,3})+(\nu_{4,4})\right]\hspace{1mm}
	\end{eqnarray}
where $\nu_{4,1}=||(A_1+A_2)|\psi\rangle_{AB}||_2,\nu_{4,2}=||(A_2+A_3)|\psi\rangle_{AB}||_2, \nu_{4,3}=||(A_3+A_4)|\psi\rangle_{AB}||_2,\nu_{4,4}=||(A_4-A_1)|\psi\rangle_{AB}||_2$.
 Hence, 
 \ba\nonumber
 [(\mathcal{
C}_{4})_{AB}]_{Q}&\leq&\sqrt{4\big[(\nu_{4,1})^2+(\nu_{4,2})^2+(\nu_{4,3})^2+(\nu_{4,4})^2\big]}\\
 &=&\sqrt{4\big[8+\{A_2,(A_1+A_3)\}+\{A_4,(A_3-A_1)\}\big]}\\
 \nonumber\ea
 Considering $A_2=(A_1+A_3)/v'_4$ and $A_4=(A_3-A_1)/v''_4$, we get 
 \ba 
 \nonumber
 [(\mathcal{
C}_{4})_{AB}]_{Q}&\leq&\sqrt{4\big[8+2\sqrt{2+\{A_1,A_3)\}}+2\sqrt{2-\{A_1,A_3)\}}\big]}\\
\nonumber
  &=&\sqrt{4\bigg[8+2\bigg(\sqrt{4+2\sqrt{4-\{A_1,A_3)\}^2}\bigg)}\bigg]}
\ea 
 The expression is  maximized when $\{A_1,A_3\}=0$ which implies $\nu'_4=\nu''_4=\sqrt{2}$ along with   $\{A_1,A_2\}=\sqrt{2},\{A_2,A_3\}=\sqrt{2}$, $\{A_1,A_4)\}=-\sqrt{2},\{A_3,A_4\}=\sqrt{2}$, $\{A_2,A_4\}=0$. Thus, we get the conditions $A_2=(A_3+A_1)/\sqrt{2}$ and $A_4=(A_3-A_1)/\sqrt{2}$. We can then explicitly find  $\nu_{4,1}=\nu_{4,2}=\nu_{4,3}=\nu_{4,4}=\sqrt{2+\sqrt{2}}$.  Hence, the optimal quantum value we get is 
 \ba[(\mathcal{
C}_{4})_{AB}]_{Q}^{opt}=4\sqrt{2+\sqrt{2}}\ea
Here, for an example, we provided the optimal value derivation for one odd and one even value of $m$.
 Following the  similar SOS approach, we can derive the optimal quantum value for  arbitrary $m$, which can be written as 
 \ba
 [(\mathcal{
C}_{m})_{AB}]_{Q}^{opt}=max\sum\limits_{i=1}^{m}\nu_{m,i}
 \ea
 where $\nu_{m,i}=||(A_i+A_{i+1})|\psi\rangle_{AB}||_2=\sqrt{2+\langle\{A_i,A_{i+1}\}\rangle}$. This will provide  $[(\mathcal{
C}_{m})_{AB}]_{Q}^{opt}=2m\cos(\pi/2m)$ which is a well known result \cite{Brau1989,Wehner2006}. The derivation of optimal quantum value $[(\mathcal{C}_{m})_{AB}]_{Q}^{opt}$ uniquely self-tests the required observables and state. We again want to add here that the optimal quantum violation of chained Bell inequality has been derived so far by considering a two-qubit entangled state. In contrast, our derivation does not assume the dimension of the system. Details of the derivation for arbitrary $m$ are quite lengthy and deserve a separate publication elsewhere. 
  
				\subsection{A new set of $n$-locality inequalities}
				\begin{figure}[h]\hspace*{-0.6cm}
				\includegraphics[scale=0.4]{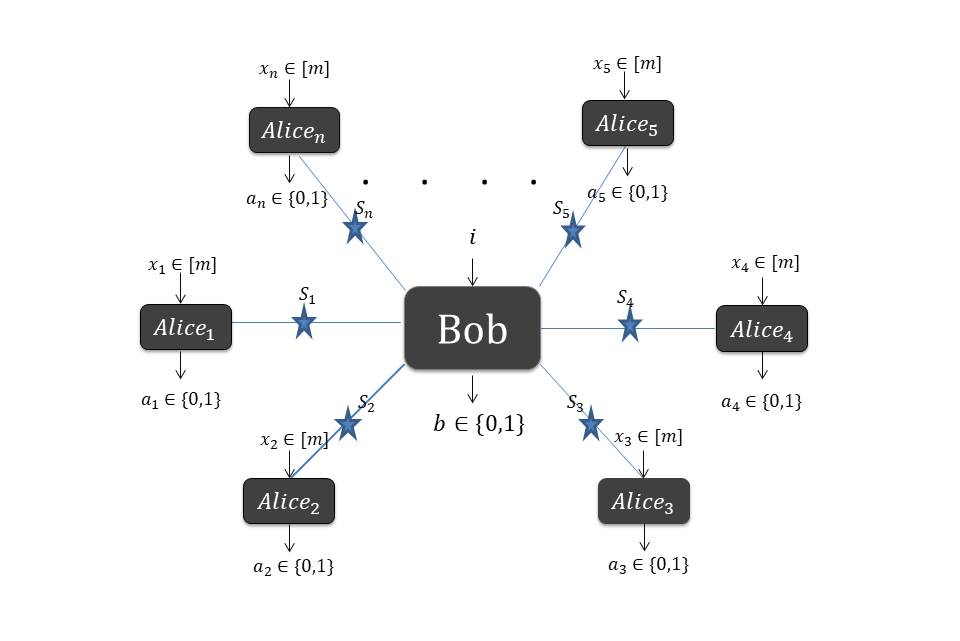}
				\caption{
				$n$-locality scenario in a star-network configuration. Here, each Alice$_k$'s input $x_k\in[m]( \forall k\in[n])$ always and  Bob's input $i\in[2^{m-1}]$  and  $i\in[m]$ represents two different family of  $n$-locality inequalities. }
				 \label{FIG. 2}
				\end{figure}
				Now,  we formulate a new set of $n$-locality inequalities.
			Consider that each party in the star-network configuration  as depicted  in Fig. \ref{FIG. 2},  performs  $m$ number of dichotomic measurements. We define a suitable expression as 
	\ba
	\Xi_m=\sum\limits_{i=1}^{m}|\mathcal{J}_{m,i}|^{1/n}
	\ea
	where $\mathcal{J}_{m,i}=\langle\prod\limits_{k=1}^{n}(A^k_{x_k}+A^k_{x_k+1})B_i\rangle$ where  $x_k=i\in[m]$ and $A^{k}_{m+1}=-A^k_1$.
	 Using the inequality in  Eq.(\ref{Tavakoli}), we can write
	\ba
\Xi_m\leq \bigg|\prod\limits_{k=1}^{n}\bigg(\sum\limits_{i=1}^{m-1}(A^k_{x_k}+A^k_{x_k+1})B_i+(A^k_{m}-A^k_{1})B_m\bigg)^{1/n}\bigg|
	\ea
	Since each obervable, $A^k_{x_k}$ and $B_i$ ( with $i\in[m]$ and  $k\in[n]$) is dichotomic,  we prove the $n$-locality inequality for arbitrary $m$ is given by  \ba\label{Cmnl}
(\Xi_m)_{nl}\leq 2m-2
	\ea

 We first demonstrate the quantum violation of the inequality in Eq.(\ref{Cmnl})  for $m=3$,  i.e.,  when each  party Alice$_k$ and Bob performs the measurements of three dichotomic observables  $A^{k}_{x_k}$ and $B_i$, where  $x_k,i\in[3]$ and $k\in[n]$. There are $n$ independent sources $S_k$, each  $S_k$ sends  an  entangled  state $\rho_{A_kB}$ to Alice$_k$ and Bob.
 The corresponding inequality is then  given by 
 \ba
 \label{Xi3}
 (\Xi_{3})_{nl}=|I_3|^{1/n}+|J_3|^{1/n}+|K_3|^{1/n}\leq 4
 \ea
 where $I_3, J_3$ and $ K_3$ are suitable linear combinations as follows:
\ba
I_3&=&\langle\prod\limits_{k=1}^{n}(A_1^k+A_2^k)B_1\rangle;
J_3=\langle\prod\limits_{k=1}^{n}(A_2^k+A_3^k)B_2\rangle;\\
K_3&=&\langle\prod\limits_{k=1}^{n}(A_3^k-A_1^k)B_3\rangle
\nonumber
\ea

  Now,  we  use SOS to optimize $\Xi_3$without assuming  the dimension of the system. Following the  method stated earlier, 
  to obtain $(\Xi_{3})^{opt}_{Q}$, let us consider that $(\Xi_{3})^{opt}_{Q}\leq \mathbb{\beta}_{3}$  where $\beta_{3}$ is clearly the upper bound of $(\Xi_{3})_{Q}$.  This is equivalent to showing that there is a positive semidefinite operator $\langle \mu_3\rangle_{Q}\geq 0$ which can be expressed as $\langle \mu_3\rangle_{Q}=-(\Xi_3)_{Q}+\beta_3$.	Defining  a set of suitable positive operators $L^n_{3,i}$ with $i=1,2,3$ which are polynomial functions of   $A^{k}_{x_{k}}$ ($x_{k}\in[3]$ and $k\in[n]$),  $B_{1},B_{2}$ and $B_{3}$, we can write
	\begin{equation}
	\label{mu3}
	\mu_3=\dfrac{(\nu^n_{3,1})^{1/n}}{2}(L^n_{3,1})^{\dagger}L^n_{3,1}+ \dfrac{(\nu^n_{3,2})^{1/n}}{2}(L^n_{3,2})^{\dagger}L^n_{3,2}+\dfrac{(\nu^n_{3,3})^{1/n}}{2}(L^n_{3,3})^{\dagger}L^n_{3,3}\hspace{5mm}
	\end{equation} Here  $\nu^n_{3,i}=\prod\limits_{k=1}^{n}(\nu^n_{3,i})_{A_{k}}$ are suitable positive numbers which  will be specified soon. The optimal quantum value of $(\Xi_{3})_{Q}$ is obtained if $\langle \mu_3\rangle_{Q}=0$, implying that 
$L^n_{3,i}|\psi\rangle=0$.
	Let us  consider  the  positive operators $L^n_{3,i}$ as
	\ba
	\label{L3}
	\nonumber
	L^n_{3,1}|\psi\rangle&=&\bigg|\prod\limits_{k=1}^{n}\left(\frac{{A}^{k}_{1}+{A}^{k}_{2}}{(\nu^n_{3,1})_{A_{k}}}
		\right)|\psi\rangle
		\bigg|^{1/n} -|B_1| |\psi\rangle|^{1/n}\\
L^n_{3,2}|\psi\rangle&=&\bigg|\prod\limits_{k=1}^{n}\left(\frac{(A_2^k+A_3^k)}{(\nu^n_{3,2})_{A_{k}}}
		\right)|\psi\rangle
		\bigg|^{1/n} -|B_2|\psi\rangle|^{1/n}\\
		\nonumber
L^n_{3,3}|\psi\rangle&=&\bigg|\prod\limits_{k=1}^{n}\left(\frac{(A_3^k-A_1^k)}{(\nu^n_{3,3})_{A_{k}}}
		\right)|\psi\rangle
		\bigg|^{1/n} -|B_3|\psi\rangle|^{1/n}
	\ea
	where $(\nu^n_{3,1})_{A_{k}}=||({A}^{k}_{1}+{A}^{k}_{2})|\psi\rangle||_{2}=\sqrt{2+\langle\{A^{k}_{1},A^{k}_{2}\}\rangle}$ and similarly for other $(\nu^n_{3,i})_{A_k}$s $ (\forall k\in[n])$. For notational convenience, we write $|\psi\rangle_{A_1A_2\ldots A_n B}=|\psi\rangle$.       
	Putting Eq.(\ref{L3}) in Eq. (\ref{mu3}), we get 	$\langle\mu_{3}\rangle_{Q}=-(\Xi_{3})         +(\nu^n_{3,1})^{1/n}+(\nu^n_{3,2})^{1/n}+(\nu^n_{3,3})^{1/n}$. 	Since $\langle\mu_{3}\rangle_{Q}\geq 0$, we have 
	\ba\label{C3Qopt1}
	(\Xi_{3})_{Q}^{opt}= max\left((\nu^n_{3,1})^{1/n}+(\nu^n_{3,2})^{1/n}+(\nu^n_{3,3})^{1/n}\right)
	\ea
	Using the inequality  Eq.(\ref{Tavakoli})
	we can write
\ba
\label{C3Qopt}
(\Xi_{3})_{Q}^{opt}&\leq& max\bigg[\prod\limits_{k=1}^{n}\bigg((\nu_{3,1})_{A_k}+(\nu_{3,2})_{A_k}+(\nu_{3,3})_{A_k}\bigg)^{1/n}\bigg]\\
\nonumber
\ea
	Now, if we consider the entangled state shared by the source $S_k$ between  Alice$_k$ and Bob, and say Bob performs measurement $B_i,i\in[3]$ on his part of subsystem  then the  relevant chained Bell inequality $(\mathcal{C}_{3})_k\leq 4$  can be optimized using the similar SOS  approach as stated above. Hence, the optimal quantum value of chained Bell expression for all $k\in[n]$ is
\ba\label{B3kQopt}[(\mathcal{C}_{3})_k]_{Q}^{ opt}=max[(\nu^n_{3,1})_{A_k}+(\nu^n_{3,2})_{A_k}+(\nu^n_{3,3})_{A_k}]\ea

From Eq.(\ref{C3Qopt}) and Eq.(\ref{B3kQopt}), we can write
	\begin{equation}
	\label{I3q}
	(\Xi_{3})^{opt}_{Q}\leq\ \prod\limits_{k=1}^{n}\left([(\mathcal{C}_{3})_{k}]^{opt}_Q\right)^{1/n}
	\end{equation}
Hence if for a given $k$ ($k\in[n]$), the state $\rho_{A_kB}$ violates relevant chained Bell inequality and each source $S_k$ shares the state $\rho_{A_kB}$ between Alice$_k$ and Bob ($\forall k\in[n]$), then $\bigotimes \limits_{k=1}^{n}\rho_{A_kB}$ violates $n$-locality inequality in  Eq.(\ref{Xi3}).  	
	
Similarly, if we consider the scenario where each party performs four dichotomic observables ($m=4$) then, the corresponding inequality will be 
\ba
 \label{I4}
 \Xi_{4}=|I_4|^{1/n}+|J_4|^{1/n}+|K_4|^{1/n}+|L_4|^{1/n}\leq 6
 \ea
 where $I_4, J_4, K_4$ and $L_4$ are suitable linear combinations as follows
\ba
I_4&=&\langle\prod\limits_{k=1}^{n}(A_1^k+A_2^k)B_1\rangle, \hspace{2mm}
J_4=\langle\prod\limits_{k=1}^{n}(A_2^k+A_3^k)B_2\rangle\\
\nonumber
K_4&=&\langle\prod\limits_{k=1}^{n}(A_3^k+A_4^k)B_3\rangle, \hspace{2mm}
L_4=\langle\prod\limits_{k=1}^{n}(A_4^1-A_1^1)B_4\rangle
\ea

Along the same line of derivation, we can show
\begin{equation}
	\label{I4q}
	(\Xi_{4})^{opt}_{Q}\leq\ \prod\limits_{k=1}^{n}\left([(\mathcal{C}_{4})_{k}]^{opt}_Q\right)^{1/n}
	\end{equation}
	Following this similar approach, in quantum theory, to obtain $(\Xi_m)^{opt}_{Q}$, we will use SOS approach again. Let us consider $(\Xi_m)^{opt}_{Q}\leq \beta_m$, where $\beta_m$ is the upper bound of $(\Xi_m)^{opt}_{Q}$.  This is equivalent to showing that there is a positive semidefinite operator $\langle \mu_m\rangle_{Q}\geq 0$ which can be expressed as $\langle \mu_m\rangle_{Q}=-(\Xi_m)_{Q}+\beta_m$.	As earlier, by invoking a set of suitable positive operators $L^n_{m,i}$ which are polynomial functions of   $A^{k}_{x_{k}}$, $B_{i}$, $(x_{k},i\in[m], k\in[n])$, we can write
	\begin{equation}
	\label{mum}
	\mu_m=\sum\limits_{i=1}^{m}\dfrac{(\nu^n_{m,i})^{1/n}}{2}(L^n_{m,i})^{\dagger}L^n_{m,i}\hspace{5mm}
	\end{equation} and $\nu^n_{m,i}=\prod\limits_{k=1}^{n}(\nu^n_{m,i})_{A_{k}}$ are suitable positive numbers. The optimal quantum value of $(\mathcal{C}_{m})_{Q}$ is obtained if $\langle \mu_m\rangle_{Q}=0$, implying that 
	\begin{align}
	\label{Lnmi}
	L^n_{m,i}|\psi\rangle=0, \forall i\in[m]
	\end{align}
We consider  a set of suitable positive operators $L^n_{m,i}$ as
	
		\ba
	\label{Lm}
	L^n_{m,i}|\psi\rangle&=&\bigg|\prod\limits_{k=1}^{n}\left(\frac{{A}^{k}_{i}+{A}^{k}_{i+1}}{(\nu^n_{m,i})_{A_{k}}}
		\right)|\psi\rangle
		\bigg|^{1/n} -|B_i |\psi\rangle|^{1/n}
		\ea
	where $(\nu^n_{m,i})_{A_{k}}=||({A}^{k}_{i}+{A}^{k}_{i+1})|\psi\rangle_{A_kB}||_{2}=\sqrt{2+\langle\{A^{k}_{i},A^{k}_{i+1}\}\rangle}$, for each $k\in[n]$.         
	Putting 	$L^n_{m,i}|\psi\rangle$ of Eq.(\ref{Lm}) in Eq. (\ref{mum}), we get 	$\langle\mu_{m}\rangle_{Q}=-(\Xi_{m})         +\sum\limits_{i=1}^{m}(\nu^n_{m,i})$. 	Since $\langle\mu_{m}\rangle_{Q}\geq 0$, we have 
	\begin{eqnarray}(\Xi_{m})_{Q}^{opt}&=& max\bigg[\sum\limits_{i=1}^{m}(\nu^n_{m,i})\bigg]
	\end{eqnarray}
	If we consider the system shared by the source $S_k$ between  Alice$_k$ and Bob, and say Bob performs measurement $B_i,i\in[m]$ on his part of the sysrem, then the  relevant chained Bell inequality for Alice$_k$ and Bob is given by 
	\ba
	(\mathcal{C}_{m})_k=\sum\limits_{i=1}^{m}(A^k_i+A^k_{i+1})B_i\leq 2m-2, \forall k\in[n]
	\ea
	with $A_{m+1}=-A_{1}$.
	To optimize $	[(\mathcal{C}_{m})_k]_Q$, using  similar SOS approach as stated earlier, we get 
\ba[(\mathcal{C}_{m})_k]_{Q}^{ opt}=max\sum\limits_{i=1}^{m}(\nu^n_{m,i})_{A_k}=2m\cos\frac{\pi}{2m}\ea	

Now, using the inequality (\ref{Tavakoli}), we get 

\ba
\nonumber
(\Xi_{m})_{Q}^{opt}&\leq& \prod\limits_{k=1}^{n}\left(\sum\limits_{i=1}^{m}(\nu^n_{m,i})_{A_k}\right)^{1/n}
\ea

	which in turn provides 
	\begin{equation}
	\label{Imq}
	({\Xi}_{m})^{opt}_{Q}\leq\ \prod\limits_{k=1}^{n}\left([(\mathcal{C}_{m})_k]^{opt}_Q\right)^{1/n}
	\end{equation}
Hence, if for each $k\in[n]$, the state $\rho_{A_kB}$ violates relevant chained Bell inequality and each source $S_k$ shares the state $\rho_{A_kB}$ between Alice$_k$ and Bob ($\forall k\in[n]$), then $\bigotimes\limits_{k=1}^n\rho_{A_kB}$ violates the $n$-locality inequality in star-network configuration  in Eq.(\ref{Cmnl}). Note again that the whole derivation is irrespective of the dimension of the system.
 
\section{Summary and Discussion}  
\label{SecVII}
In summary, we characterized network nonlocality in generalized star-network configuration through the quantum violation of various forms of $n$-locality inequalities. All the previous works demonstrated the network nonlocality by assuming that each source emits a two-qubit entangled state and cannot provide the device-independent certifications. In this work, we have gone beyond the dimensional restriction and provided the network nonlocality without assuming the dimension of the quantum system, and each party performs black-box measurements. Hence the optimal quantum violation of a given $n$-locality inequality enables a device-independent self-testing scheme. We used a variant of the known SOS approach for deriving the optimal quantum violations of $n$-locality inequalities, which plays a crucial role in this work.

We first considered the simplest bilocal scenario in \cite{Cyril2012,Bran2010} and reproduced the results of Gisin \emph {et. al}.\cite{Gisin2017}. In \cite{Gisin2017} it has been demonstrated that if entangled states originating from two independent sources violate CHSH inequality, then the joint entangled state is capable of exhibiting the violation of bilocality inequality. However, we note again that the work in \cite{Gisin2017} assumes the two-qubit entangled state and derivation is based on the famous work of Horodeckis\cite{Horodecki2009}. In this work, we demonstrated Gisin \emph {et. al}.\cite{Gisin2017} result without imposing any restriction to the dimension, and measurements are also taken to be uncharacterized. We further generalized our treatment to demonstrate the above correspondence between the violations of CHSH inequality and $n$-locality inequality in a star-network topology.

We provided further generalization by considering an arbitrary number of inputs for every party in the star-network scenario. In particular, we proposed two different families of $n$-locality inequalities and demonstrated the correspondence mentioned above between their violation with the violation of relevant Bell inequalities. Notably, the derivation of optimal quantum violation of any inequality is irrespective of the dimension of the system. In an arbitrary input scenario, we first considered a star-network configuration where each of the $n$ edge parties receives $m$ inputs, and the central party received $2^{m-1}$ inputs. A family of $n$-locality inequalities in this scenario was formulated in \cite{Sneha2021}. We characterized the quantum violation of those inequalities and showed correspondence between that quantum violation of relevant bipartite inequalities in \cite{Ghorai2018} and $n$-locality inequalities. We then proposed another family of $n$-locality inequalities in star-network configuration where each of $n$ edge party and the central party receive $m$ inputs. We demonstrated a similar correspondence between the quantum violation of $n$-locality inequalities and chained Bell inequalities. We note again that throughout this paper, we assumed the measurement devices are uncharacterized and imposed no constraints on the dimension of the system. The optimal quantum violations of various $n$-locality inequalities derived in this paper enable device-independent self-testing of states and measurements. 
 
We conclude by stating a few open questions. Studying non-$n$-locality in various other topologies of quantum networks for arbitrary $m$ could be an interesting line of future work. Note that multiple sources in a network open up the possibility of various forms of attacks from eavesdroppers. It is then relevant to ask whether the violation of our generalized $n$-locality inequalities can prevent such attacks so that the non-$n$-locality can be useful for device-independent information processing. In a standard device-independent test, one assumes that the adversary Eve supplies the sources and devices to untrusted parties. It may then be possible that the systems produced by the independent sources are correlated with Eve's system, and by performing suitable measurements, Eve could gain information about the outcomes of Alices and Bob. It is worth noting that a set of local correlations is convex, but due to the nonlinear $n$-locality condition, a mixture of $n$-local correlations is not necessarily $n$-local, i.e., the set of $n$-local correlations is not convex \cite{Cyril2012}. Therefore the approaches establishing the standard Bell nonlocality as a device-independent resource for quantum information processing are not applicable for network Bell experiments. This could be another interesting direction of study.

Recently, an interesting scheme has been developed \cite{Lee2018} to demonstrate how the quantum violation of $n$-locality inequalities allows device-independence, secure against Eve's attacks. Following the same line, it would then be interesting to study the device-independence nature of our various $n$-locality inequalities by considering various forms of attacks. We also note that the standard Bell scenario involving many inputs has been proven beneficial for device-independent information processing compared to the protocols involving only two inputs. Masanes \emph{et al.}\cite{masa11} demonstrated that using chain Bell inequalities \cite{Brau1989} other than two-input Bell-CHSH can provide better key rates in the presence of noise. In a recent work \cite{cab21}, the authors demonstrated that a device-independent quantum key distribution protocol based on a Bell's inequality with more than two inputs works better for lower detection efficiency compared to the protocols with two inputs. It would then be interesting to examine whether our $n$-locality inequalities for an arbitrary $m$ number of inputs per Alice minimizes Eve's information about the outcomes of Alice and Bob compared to the case of two inputs case. Studies towards this line could lead to an exciting avenue for future research and calls for further investigation.

\end{document}